\documentclass[letterpaper]{article} 
\usepackage{aaai25}  
\usepackage{times}  
\usepackage{helvet}  
\usepackage{courier}  
\usepackage[hyphens]{url}  
\usepackage{graphicx} 
\usepackage{natbib}  
\usepackage{caption} 
\usepackage[utf8]{inputenc} 
\usepackage[T1]{fontenc}    
\usepackage{url}            
\usepackage{booktabs}       
\usepackage{amsfonts}       
\usepackage{nicefrac}       
\usepackage{microtype}      
\usepackage[table]{xcolor}         
\usepackage{xcolor}         
\usepackage{graphicx}
\usepackage{subfigure}
\usepackage{amsmath}
\usepackage{multirow}
\usepackage{multicol}
\usepackage{bbding}

\usepackage{algorithm}
\usepackage{algorithmic}

\usepackage{amssymb}
\usepackage{footnote}
\usepackage{url}
\usepackage{enumitem}
\usepackage{algorithm}
\usepackage{algorithmic}
\usepackage{listings}
\usepackage{tabularx}
\usepackage {pifont} 
\usepackage{bm}
\usepackage{tablefootnote}
\usepackage{cutwin}
\usepackage{colortbl}
\usepackage[table]{xcolor}
\definecolor{cvprblue}{rgb}{0.21,0.49,0.74}
\definecolor{lightgray}{gray}{0.9}
\definecolor{linecolor}{rgb}{0.82, 0.94, 0.75}

\usepackage{amsmath,amsfonts,bm}

\def\eqref#1{equation~\ref{#1}}

\def\1{\bm{1}}

\def\mI{{\bm{I}}}

\def\mX{{\bm{X}}}
\def\mY{{\bm{Y}}}
\def\mZ{{\bm{Z}}}

\DeclareMathAlphabet{\mathsfit}{\encodingdefault}{\sfdefault}{m}{sl}
\SetMathAlphabet{\mathsfit}{bold}{\encodingdefault}{\sfdefault}{bx}{n}

\newcommand*{\op}[1]{\operatorname{#1}}

\usepackage{newfloat}
\usepackage{listings}
\DeclareCaptionStyle{ruled}{labelfont=normalfont,labelsep=colon,strut=off} 
\floatstyle{ruled}
\newfloat{listing}{tb}{lst}{}
\floatname{listing}{Listing}
\newcommand{\unet}{{{U-Net}}} 
\newcommand{\model}{{{U-KAN}}}  

\title{U-KAN Makes Strong Backbone for Medical Image Segmentation and Generation}

\author{
    Chenxin Li\textsuperscript{\rm 1}\equalcontrib,
    Xinyu Liu\textsuperscript{\rm 1}\equalcontrib,
   Wuyang Li\textsuperscript{\rm 1}\equalcontrib,
   Cheng Wang\textsuperscript{\rm 1}\equalcontrib,\\
    Hengyu Liu\textsuperscript{\rm 1},
    Yifan Liu\textsuperscript{\rm 1},
    Zhen Chen\textsuperscript{\rm 2},
    Yixuan Yuan\textsuperscript{\rm 1},
}
\affiliations{
    \textsuperscript{\rm 1}The Chinese University of Hong Kong, \textsuperscript{\rm 2}CAIR, HKISI-CAS\\

}

\usepackage{bibentry}

\begin{document}

\maketitle

\begin{abstract}
\unet~has become a cornerstone in various visual applications such as image segmentation and diffusion probability models.
While numerous innovative designs and improvements have been introduced by incorporating transformers or MLPs, the networks are still limited to linearly modeling patterns as well as the deficient interpretability. 
To address these challenges, our intuition is inspired by the impressive results of the Kolmogorov-Arnold Networks (KANs) in terms of accuracy and interpretability, which reshape the neural network learning via the stack of non-linear learnable activation functions derived from the Kolmogorov-Anold representation theorem.
Specifically, in this paper, we explore the untapped potential of KANs in improving backbones for vision tasks.
We investigate, modify and re-design the established U-Net pipeline by integrating the dedicated KAN layers on the tokenized intermediate representation, termed \model.
Rigorous medical image segmentation benchmarks verify the superiority of \model~by higher accuracy even with less computation cost. 
We further delved into the potential of \model~as an alternative U-Net noise predictor in diffusion models, demonstrating its applicability in generating task-oriented model architectures.
Project page:  \url{https://yes-u-kan.github.io/}.

\end{abstract}

%

\section{Introduction}

Over the past decade, numerous works have focused on developing efficient and robust segmentation methods for medical imaging~\cite{shen2017deep,sun2022few,li2022hierarchical,li2021consistent}, driven by the need for computer-aided diagnosis and image-guided surgical systems~\cite{liu2024endogaussian,liu2024lgs, li2024endosparse, liu2022source, liu2021consolidated,ali2024assessing}. Among these, U-Net \cite{ronneberger2015u} is a landmark work that initially demonstrated the effectiveness of encoder-decoder convolutional networks with skip connections for medical image segmentation~\cite{wang2022medical,li2021unsupervised,ding2022unsupervised,xu2022afsc}, and has also shown promising results in many image translation tasks~\cite{torbunov2023uvcgan,kalantar2021ct}. Additionally, recent diffusion models have utilized U-Net, training it to iteratively predict the noise to be removed in each denoising step~\cite{ho2020ddpm,rombach2022ldm,saharia2022imagen}.

Since the inception of U-Net~\cite{ronneberger2015u}, a series of crucial modifications have been introduced, especially in the subfield of medical imaging, including U-Net++\cite{zhou2018unet++}, 3D U-Net\cite{ccccek20163d}, V-Net~\cite{milletari2016v}, and Y-Net~\cite{mehta2018net}. U-NeXt~\cite{valanarasu2022unext} and Rolling U-Net\cite{liu2024rolling} integrate hybrid approaches involving convolutional operations and MLP to optimize the efficacy of segmentation networks, enabling their deployment at point-of-care settings with limited resources.
Recently, numerous transformer-based networks have been utilized to enhance the U-Net backbone for medical image segmentation. These networks have demonstrated effectiveness in addressing global context and long-range dependencies\cite{raghu2021vision,hatamizadeh2023global,li2022sigma,li2023novel}. Examples include Trans-UNet~\cite{chen2021transunet}, which adopts ViT architecture\cite{dosovitskiy2020image} for 2D medical image segmentation using U-Net, and other transformer-based networks like MedT~\cite{valanarasu2021medical} and UNETR\cite{hatamizadeh2022unetr}. Although showing great scaling capacity due to the sophisticated designs, transformers tend to overfit when dealing with limited datasets, indicating their data-hungry nature \cite{deit, liu2023efficientvit}. In contrast, structured state-space sequence models (SSMs)~\cite{fu2022hungry,peng2023rwkv,gu2023mamba} 
have recently shown high efficiency and effectiveness in long-sequence modeling. For medical image segmentation, U-Mamba~\cite{ma2024u} and SegMamba~\cite{xing2024segmamba} have proposed task-specific architectures with Mamba blocks respectively based on nn-UNet~\cite{isensee2021nnu} and Swin UNETR~\cite{hatamizadeh2021swin}, achieving promising results in various vision tasks.

{While existing U-shape variations have been advanced in fine-trained medical scenarios, e.g., medical image segmentation, they still have fundamental challenges due to their sub-optimal kernel design and the unexplainable nature. Concretely, first, they typically employ conventional kernels\footnote{Such operations include convolution, Transformers, and MLPs, etc.} to capture the spatial dependence between local pixels, which are {limited to linearly modeling patterns} and relationships across different channels in latent space. This makes it challenging to capture complex nonlinear patterns.} 
Such intricate nonlinear patterns among channels are prevalent in visual tasks, such as medical imaging, where images often have intricate diagnostic characteristics. This complexity implies that feature channels might possess varying clinical relevance, representing different anatomical components or pathological indicators. {Second, they mostly conduct empirical network search and heuristic model design to find the optimal architecture, {ignoring the interpretability and explainability} in existing black-box U-shape models. In existing U-shape variations, this unexplainable property poses a significant risk in clinical decision-making, further preventing the truth-worth of diagnostic system design. Recently, Kolmogorov-Arnold Networks (KANs) have attempted to open the black box of conventional network structures with superior interpretability, revealing the great potential of white-box network reseach~\cite{yu2024white,pai2024masked}.} Considering the excellent architecture properties merged in KANs, it makes sense to effectively leverage KAN to bridge the gap between the network's physical attributes and empirical performance. 

In this endeavor, we have embarked on the exploration of a universally applicable U-KAN framework, denoted as U-KAN, marking an inaugural attempt to integrate advanced KAN into the pivotal visual backbone of UNet, through a convolutional KAN mixed architectural style. Notably, adhering to the benchmark setup of U-Net, we employ a multilayered deep encoder-decoder architecture with skip connections, incorporating a novel tokenized KAN block at higher-level representations proximate to the bottleneck. This block projects intermediate features into tokens, subsequently applying the KAN operator to extricate informative patterns. The proposed U-KAN benefits from the alluring attributes of KAN networks in terms of non-linear modeling capabilities and interpretability, distinguishing it prominently within the prevalent U-Net architecture. Empirical evaluations on stringent medical segmentation benchmarks, both quantitative and qualitative, underscore U-KAN's superior performance, outpacing established U-Net backbones with enhanced accuracy even with lower computation cost. Our investigation further delves into the potentiality of \model~as an alternative U-Net noise predictor in diffusion models, substantiating its relevance in generating task-oriented model architectures. 
In a nutshell, U-KAN signifies a steady step toward the design that incorporates mathematics theory-inspired operators into efficient visual pipelines and foretells its prospects in extensive visual applications.
Our contributions can be summarized as follows:
\begin{itemize}
    \item We present the first effort to incoporate the advantage of emerging KAN, improving the established U-Net pipeline to be more accurate, efficient, and interpretable.
    \item We propose a tokenized KAN block to effectively steer the KAN operators to be compatible with the existing convolution-based designs.
    \item We empirically validate \model~on a wide range of medical segmentation benchmarks, achieving impressive accuracy and efficiency.
    \item The application of \model~to existing diffusion models as an improved noise predictor demonstrates its potential in backbone generative tasks and broader vision settings.
\end{itemize}

\section{Related Work}
\subsection{
U-Net Backbone for Medical Image Segmentation
}
Medical image segmentation~\cite{ronneberger2015u,segresnet,li2024u,li2022domain} is a challenging task to which deep learning methods have been extensively applied and achieved breakthrough advancements in recent years~\cite{shen2017deep,liu2024lgs,li2024endosparse,yang2023mrm,liu2023decoupled,li2021htd, chen2023medical,liu2022intervention,wuyang2021joint}. U-Net \cite{ronneberger2015u} is a popular network structure for medical image segmentation. Its encoder-decoder architecture effectively captures image features. The CE-Net \cite{gu2019ce} further integrates a contextual information encoding module, enhancing the model's receptive field and semantic representation capabilities. Unet++ \cite{zhou2018unet++} proposes a nested U-Net structure that fuses multi-scale features to improve segmentation accuracy.
In addition to convolution-based methods, Transformer-based models have also gained attention. The Vision Transformer \cite{dosovitskiy2020image} demonstrates the effectiveness of Transformers in image recognition tasks. The Medical Transformer \cite{valanarasu2021medical} and TransUNet \cite{chen2021transunet} further incorporate Transformers into medical image segmentation, achieving satisfying performance.
Moreover, techniques such as attention mechanism \cite{schlemper2019attention} and multi-scale feature fusion \cite{huang2020unet} are widely used in medical image segmentation tasks. 3D segmentation models like Multi-dimensional Gated Recurrent Units \cite{andermatt2016multi} and Efficient Multi-Scale 3D CNN \cite{kamnitsas2017efficient} also yield commendable results.
In summary, medical image segmentation is an active research field where deep learning methods have made significant progress. 
Recently, Mamba~\cite{gu2023mamba} has achieved a groundbreaking milestone with its linear-time inference and efficient training process by integrating selection mechanism and hardware-aware algorithms into previous works~\cite{gu2022parameterization, gupta2022diagonal, mehta2022long}. Building on the success of Mamba
for visual application, Vision Mamba~\cite{liu2024vmamba} and VMamba~\cite{zhu2024vision} use bidirectional Vim Block and the Cross-Scan Module, respectively, to gain data-dependent global visual context. At the same time, U-Mamba~\cite{ma2024u} and other works~\cite{xing2024segmamba, ruan2024vm} show superior performance in medical image segmentation.
As Kolmogorov–Arnold Network (KAN)~\cite{liu2024kan} has been emerged as a promising alternative for MLP and demonstrates its precision, efficiency, and interpretability, we believe now is the right time to open up the exploration of its broader applications in vision backbones.

\subsection{
U-Net Diffusion Backbone for Image Generation
}
Diffusion Probability Models, a frontier category of generative models, have emerged as a focal point in the research domain, particularly in tasks related to computer vision~\cite{ho2020ddpm,rombach2022ldm,ramesh2022dalle2}.
Unlike other categories of generative models~\cite{kingma2013vae,wang2021rethinking,goodfellow2014gan,mirza2014cgan,brock2018biggan,karras2018pggan}, 
such as Variational Autoencoders (VAE)~\cite{kingma2013vae}, Generative Adversarial Networks (GANs)~\cite{goodfellow2014gan,brock2018biggan,karras2018pggan,zhang2021generator},
and vector quantization methods~\cite{van2017vqvae,esser2021vqgan}, diffusion models introduce a novel generative paradigm. These models employ a fixed Markov chain to map the latent space, fostering complex mappings that capture the intricate structure inherent in datasets. Recently, their impressive generative prowess, from high-level detail to diversity in generated samples, has propelled breakthrough progress in various computer vision applications, such as image synthesis~\cite{ho2020ddpm,rombach2022ldm,saharia2022imagen}, image editing~\cite{avrahami2022blended,choi2021ilvr,meng2021sdedit,li2024feature},
image-to-image translation~\cite{choi2021ilvr,saharia2022palette,wang2022pretraining,li2024mapping}, and video generation~\cite{hong2022cogvideo,blattmann2023videoldm,he2022lvdm,li2024endora}.
Diffusion models consist of a \textit{diffusion process} and a \textit{denoising process}. In the diffusion process, Gaussian noise is gradually added to the input data, eventually corroding it to approximate pure Gaussian noise. In the \textit{denoising process}, the original input data is recovered from its noisy state through a learned sequence of inverse diffusion operations. Typically, convolutional U-Nets~\cite{ronneberger2015u}, the de-facto choice of backbone architecture, are trained to iteratively predict the noise to be removed at each denoising step.
Diverging from previous work that focuses on utilizing pre-trained diffusion U-Nets for downstream applications, recent work has committed to exploring the intrinsic features and structural properties of diffusion U-Nets. Free-U investigates strategically reassessing the contribution of U-Net's skip connections and backbone feature maps to leverage the strengths of the two components of the U-Net architecture. RINs~\cite{jabri2022scalable} introduced a novel, efficient architecture based on attention for DDPMs. DiT~\cite{peebles2023scalable} proposed the combination of pure transformer with diffusion, showcasing its scalable nature. In this paper, we demonstrate the potential of a backbone scheme integrating U-Net and KAN for generation, pushing the boundaries and options for generation backbone.

\subsection{Kolmogorov–Arnold Networks (KANs)
}
The Kolmogorov-Arnold theorem \cite{kolmogorov1957representation} postulates that any continuous function can be expressed as a composition of continuous unary functions of finite variables, providing a theoretical basis for the construction of universal neural network models. This was further substantiated by Hornik et al. \cite{hornik1989multilayer}, who demonstrated that feed-forward neural networks possess universal approximation capabilities, paving the way for the development of deep learning.
Drawing from the Kolmogorov-Arnold theorem, scholars proposed a novel neural network architecture known as Kolmogorov-Arnold Networks (KANs) \cite{huang2014deep}. KANs consist of a series of concatenated Kolmogorov-Arnold layers, each containing a set of learnable one-dimensional activation functions. This network structure has proven effective in approximating high-dimensional complex functions, demonstrating robust performance across various applications.
KANs are characterized by strong theoretical interpretability and explainability. Huang et al. \cite{huang2017towards} analyzed the optimization characteristics and convergence of KANs, validating their excellent approximation capacity and generalization performance. Liang et al. \cite{liang2018deep} further introduced a deep KAN model and applied it to tasks such as image classification. Xing et al. \cite{xing2018kolmogorov} deployed KANs for time series prediction and control problems.
Despite these advancements, there has been a lack of practical implementations to broadly incorporate the novel neural network model of KAN, which has strong theoretical foundations, into general-purpose vision networks. In contrast, this paper undertakes an initial exploration, attempting to design a universal visual network architecture that integrates KAN and validates it on a wide range of segmentation and generative tasks.

\section{Method}

\begin{figure*}[!t]
\begin{center}
    \includegraphics[width=0.95\linewidth]{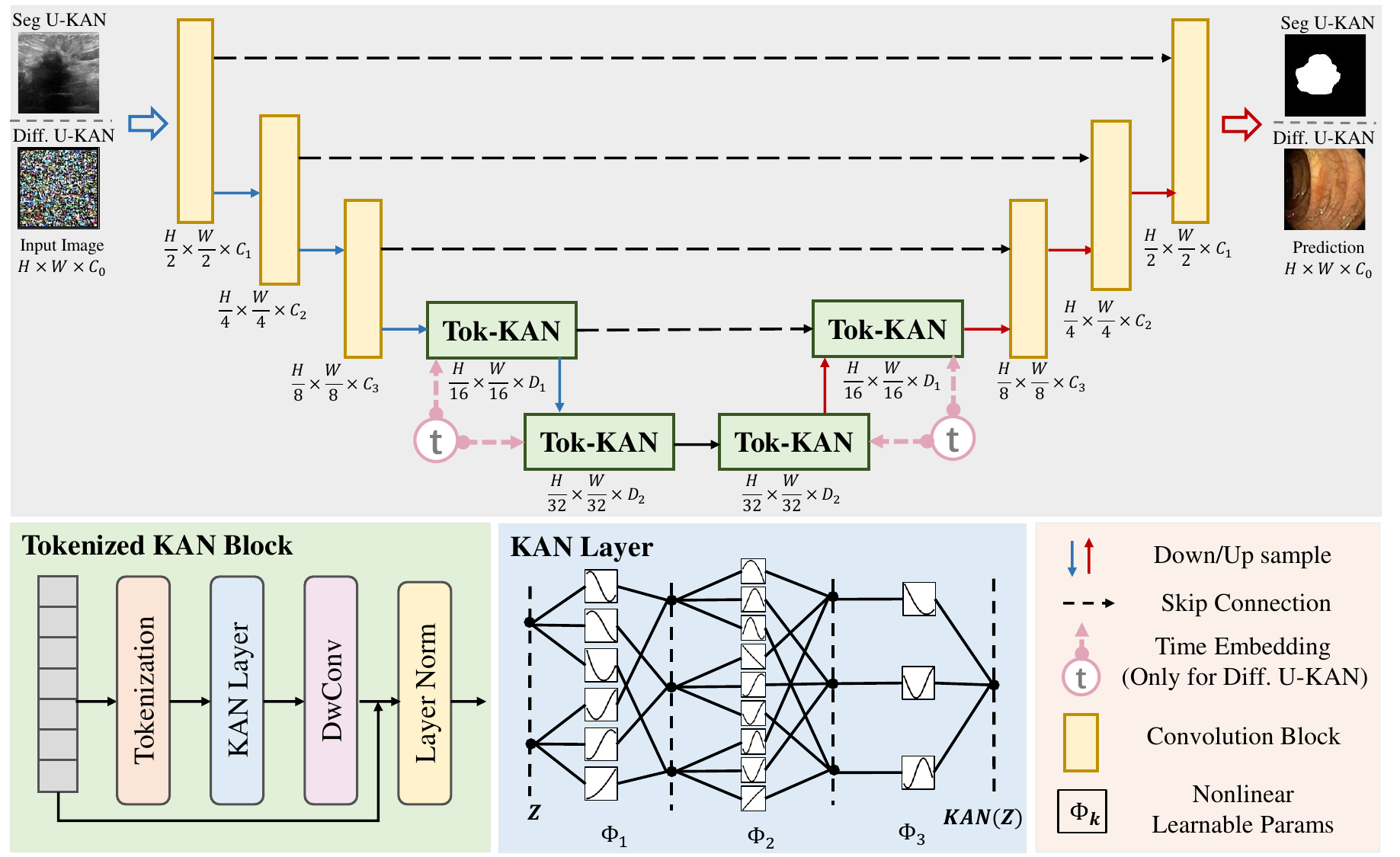}
\end{center}
\caption{
 Overview of \model~pipeline. 
     After feature extraction by several convolution blocks in Convolution Phrase, the intermediate maps are tokenized and processed by stacked Tok-KAN blocks in Tokenized KAN Phrase.
 The time embedding is only injected into the KAN blocks when applied for Diffusion U-KAN.
 }
\label{fig:method}
\end{figure*}

\paragraph{Overview}
Fig.~\ref{fig:method} illustrates the overall architecture of the proposed \model, following a two-phase encoder-decoder architecture comprising a Convolution Phrase and a Tokenized Kolmogorov–Arnold Network (Tok-KAN) Phrase. The input image traverses the encoder, where the initial three blocks utilize convolution operations, followed by two tokenized MLP blocks. The decoder comprises two tokenized KAN blocks followed by three convolution blocks. Each encoder block halves the feature resolution, while each decoder block doubles it. Additionally, skip connections are integrated between the encoder and decoder. The channel count for each block in Convolution Phrase and Tok-KAN Phrase is respectively determined by hyperparameters as $C_1$ to $C_3$ and $D_1$ to $D_3$.

\subsection{KAN as Efficient Embedder}
This research aims to incorporate Kolmogorov–Arnold Networks (KANs) into the U-Net framework. The basis of this approach is the proven high efficiency and interpretability of KANs as outlined in \cite{liu2024kan}.
A Multi-Layer Perceptron (MLP) comprising $K$ layers can be described as an interplay of transformation matrices ${W}$ and activation functions $\sigma$. This can be mathematically expressed as:
\begin{small}
\begin{equation}
\operatorname{MLP}(\mathbf{Z}) = \left({W}_{K-1} \circ \sigma \circ {W}_{K-2} \circ \sigma \circ \cdots \circ {W}_1 \circ \sigma \circ {W}_0\right) \mathbf{Z},
\end{equation}
\end{small}
where it strives to mimic complex functional mappings through a sequence of nonlinear transformations over multiple layers. Despite its potential, the inherent obscurity within this structure significantly hampers the model's interpretability, thus posing considerable challenges to intuitively understanding the underlying decision-making mechanisms.
In an effort to mitigate the issues of low parameter efficiency and limited interpretability inherent in MLPs, Liu \emph{et al.}~\cite{liu2024kan} proposed the Kolmogorov-Arnold Network (KAN), drawing inspiration from the Kolmogorov-Arnold representation theorem ~\cite{kolmogorov1961representation}. 

Similar to an MLP, a $K$-layer KAN can be characterized as a nesting of multiple KAN layers:
\begin{equation}
    \operatorname{KAN}(\mathbf{Z})=\left(\boldsymbol{\Phi}_{K-1} \circ \boldsymbol{\Phi}_{K-2} \circ \cdots \circ \boldsymbol{\Phi}_{1} \circ \boldsymbol{\Phi}_{0}\right) \mathbf{Z},
\end{equation}
where $\boldsymbol{\Phi}_i$ signifies the $i$-th layer of the entire KAN network. Each KAN layer, with $n_{in}$ -dimensional input and $n_{out}$ -dimensional output,  $\boldsymbol{\Phi}$ comprises $ n_{in} \times n_{out}$ learnable activation functions $\phi$:
\begin{equation}
    \boldsymbol{\Phi}=\left\{\phi_{q, p}\right\}, \quad p=1,2, \cdots, n_{\text {in }}, \quad q=1,2 \cdots, n_{\text {out }},
\end{equation}

The computation result of the KAN network from layer $k$ to layer $k+1$ can be expressed in matrix form $\mathbf{Z}_{k+1} = \boldsymbol{\Phi}_{k}\mathbf{Z}_{k}$, where:
\begin{equation}
\boldsymbol{\Phi}_{k} = \left(\begin{array}{cccc}
\phi_{k, 1,1}(\cdot) & \phi_{k, 1,2}(\cdot) & \cdots & \phi_{k, 1, n_{k}}(\cdot) \\
\phi_{k, 2,1}(\cdot) & \phi_{k, 2,2}(\cdot) & \cdots & \phi_{k, 2, n_{k}}(\cdot) \\
\vdots & \vdots & & \vdots \\
\phi_{k, n_{k+1}, 1}(\cdot) & \phi_{k, n_{k+1}, 2}(\cdot) & \cdots & \phi_{k, n_{k+1}, n_{k}}(\cdot)
\end{array}\right)
\end{equation}
In conclusion, KANs differentiate themselves from traditional MLPs by using learnable activation functions on the edges and parametrized activation functions as weights, eliminating the need for linear weight matrices. This design allows KANs to achieve comparable or superior performance with smaller model sizes. Moreover, their structure enhances model interpretability without compromising performance, making them suitable for various applications.

\subsection{U-KAN Architecture}

\subsubsection{Convolution Phrase}
Each convolution block is constructed of the components as follows: a convolutional layer (Conv), a batch normalization layer (BN), and a ReLU activation function. We apply a kernel size of 3x3, a stride length of 1, and a padding quantity of 1. The convolution blocks within the encoder integrate a max-pooling layer with a size of 2x2.
Formally, given an image $\mX_0=\mI \in \mathbb{R}^{H_0 \times W_0 \times C_0}$, the output of each convolution block can be elaborated as:
\begin{equation}
    \mX_{\ell} =  \op{Pool}\big(\op{Conv}(\mX_{\ell-1})\big),
\end{equation}
where $\mX_{\ell} \in  \mathbb{R}^{H_\ell \times W_\ell \times C_\ell}$ represents the output feature maps at  $\ell$-th layer. 
Given the configuration that there are $L$ blocks in the Convolution Phrase, the final output is derived as $\mX_{L}$.

\subsubsection{Tokenized KAN Phrase}

\paragraph{Tokenization}
In the tokenized KAN block, we first perform tokenization~\cite{dosovitskiy2020image,chen2024tokenunify} by reshaping the output feature of convolution phrase $\mX_L$ into a sequence of flattened 2D patches \{$\mX^i_L \in \mathbb{R}^{P^2 \cdot C_L}|i=1,..,N\}$, where each patch is of size $P \times P$ and $N=\frac{H_L\times W_L}{P^2}$ is the number of feature patches.
We then map the vectorized patches
into a latent $D$-dimensional embedding space using a trainable linear projection $\bm{\mathrm{E}} \in \mathbb{R}^{(P^2 \cdot C_L) \times D}$, as:
\begin{equation}
    \mZ_0 = [\mX^1_L \bm{\mathrm{E}}; \, \mX^2_L \bm{\mathrm{E}}; \cdots; \, \mX^{N}_L \bm{\mathrm{E}} ], \label{eq:embedding} 
\end{equation}
The linear projection $\bm{\mathrm{E}} \in \mathbb{R}^{(P^2 \cdot C_L) \times D}$ is implemented by a convolution layer with a  kernel size of 3, as 
it is shown in \cite{xie2021segformer} that a convolution layer in is enough to encode the positional information and it actually performs better than the standard positional encoding techniques. Positional encoding techniques like the ones in ViT need to be interpolated when the test and training  resolutions are not the same often leading to reduced performance.

\paragraph{Embedding by KAN Layer}
Given the obtained tokens, 
we pass them into 
a series of KAN layers ($N=3$). Followed each KAN layers, the features are passed through a efficient depth-wise convolutional layer (DwConv)~\cite{cao2022conv}  and a bacth normalization layer (BN) and a ReLU activation.
We use a residual connection here and add the original tokens as residuals. We then apply a layer normalization  (LN)~\cite{ba2016layer} and pass the output features to the next block.
Formally, the output of $k$-th Tokenized KAN block can be elaborated as:
\begin{equation}
    \mZ_k = \op{LN}(\mZ_{k-1} + \op{DwConv}(\op{KAN}(\mZ_{k-1}))),
    \label{eq:embedding} 
\end{equation}
where $\mZ_{k} \in  \mathbb{R}^{H_k \times W_k \times D_k}$ is the output feature maps at $k$-th layer. 
Given the setup that there are $K$ blocks in the Tokenized KAN Phrase, the final output is derived as $\mZ_{K}$.
In our implementation, we set $L=3$ and $K=2$.

\subsubsection{\model~Decoder}
{We follow the commonly used U-shaped architecture with dense skip connections to construct \model.} U-Net and its variations have demonstrated remarkable efficiency in medical image segmentation tasks~\cite{yang2024unicompress,li2022knowledge,xu2024immunotherapy}. This architecture leverages skip connections for the recovery of low-level details and employs an encoder-decoder structure for high-level information extraction.

Given skip-connected feature $\mZ_k$ from layer-$k$ in KAN Phrase and feature $\mZ'_{k+1}$ from the last up-sample block, the output feature $\mZ'_k$ of $k$-th up-sample block is:
\begin{equation}
    \mZ'_k = \op{Cat}\big(\mZ'_{k+1}, (\mZ_k)\big),
\end{equation}
where $\op{Cat}(\cdot)$ denotes the feature concatenation operation. Likewise,  given skip-connected feature $\mX_\ell$ from layer-$\ell$ in Convolution Phrase and feature $\mX'_{\ell+1}$ from the last up-sample block, the output feature $\mX'_\ell$ of $\ell$-th up-sample block is:
\begin{equation}
    \mX'_\ell = \op{Cat}\big(\mX'_{\ell+1}, (\mX_\ell)\big),
\end{equation}
In the context of {semantic segmentation tasks}, the final segmentation map can be derived from the output feature maps $X'_0 \in \mathbb{R}^{H_0 \times W_0 \times C_{Y}} $ at layer-$0$, where  $C_{Y}$ is the number of semantic categories and $\mY$ denotes the ground-truth segmentation and.
As a result, the segmentation loss can be:
\begin{equation}
   \mathcal{L}_{\text{Seg}} = CE\big(\mY,\text{U-KAN}(\mI)\big).
\end{equation}
where $CE$ denotes the pixel-wise cross-entropy loss.

\subsection{Extending U-KAN to Diffusion Models}
The above discussion focuses on generating segmentation masks 
given input image $\mI$ through the U-KAN.
In this section, we further extend U-KAN to a diffusion version, coined Diffusion U-KAN, which unleashes the generative capacity of KANs. Following Denosing Diffusion Probabilistic Models (DDPM)~\cite{ho2020ddpm}, Diffusion U-KAN is able to generate an image from a random Gaussian noise $\epsilon\sim\mathcal{N}(0,1)$ by gradually removing the noise. This process can be achieved by predicting the noise given a noisy input: $\epsilon_{t} = \text{U-KAN}(\mI_t, t)$, where $\mI_t$ is image $I$ corrupted by Gaussian noise $\epsilon_{t}$, $t=[1,T], T=1000$ is the time-step controlling the noise intensity, and $\mI_T\sim \mathcal{N}(0,1)$.

To this end, we conduct two modifications based on the Segmentation U-KAN to lift it to the diffusion version. First, different from only propagating features among different hidden layers, we inject learnable time embedding into each block to enable the network time-aware (see the dashed-line ``Time Embedding'' in Fig~\ref{fig:method}) and remove the DwConv and residual connections, thereby changing Eq.~\ref{eq:embedding} into the following format for the goal of generative tasks:
\begin{equation}
    \mZ_k = \op{LN}(\op{KAN}(\mZ_{k-1})) + \mathcal{F}(
    \operatorname{TE}(t)
    ),
    \label{eq:embedding_gen} 
\end{equation}
where $\mathcal{F}$ is the linear projection, $\operatorname{TE}(t)$ indicates the
time embedding for the given time step $t$~\cite{ho2020ddpm}.
Second, we modify the predicted objective to enable diffusion-based image generation. Instead of predicting segmented masks given images, Diffusion U-KAN aims to predict the noise $\epsilon_t$ given the noise-corrupted image $I_t$ and a random time-step $t=\text{Uniform}(1,T)$, which is optimized via MSE loss as follows:
\begin{equation}
   \mathcal{L}_{\text{Diff}} = || \epsilon_t - \text{U-KAN}(\mI_t, t)||_2.
\end{equation}
After optimization via the above loss function, the DDPM sampling algorithm~\cite{ho2020ddpm} is used to generate images, which leverages the well-trained Diffusion U-KAN for denoising.

\section{Experiments}

\subsection{Datasets}
We conducted a thorough evaluation of our proposed method on three distinct and heterogeneous datasets, each exhibiting unique characteristics, varying data sizes, and disparate image resolutions. These datasets are commonly utilized for tasks such as image segmentation and generation, providing a robust testing ground for the efficacy and adaptability of our method.

\paragraph{BUSI}
The BUSI dataset~\cite{al2020dataset} is made up of ultrasound images depicting normal, benign, and malignant breast cancer cases along with their corresponding segmentation maps. For our study, we utilized 647 ultrasound images representing both benign and malignant breast tumors. All these images were consistently resized to the dimensions of $256 \times 256$. The dataset offers a comprehensive collection of images that aid in the detection and differentiation of various types of breast tumors, providing valuable insights for medical professionals and researchers.

\paragraph{GlaS}
The GlaS dataset~\cite{valanarasu2021medical} is comprised of 612 Standard Definition (SD) frames from 31 sequences. Each frame possesses a resolution of $384 \times 288$ and was collected from 23 patients. This dataset is associated with the Hospital Clinic located in Barcelona, Spain. The sequences within this dataset were recorded using devices such as Olympus Q160AL and Q165L, coupled with an Extra II video processor. Following common practice~\cite{liu2024rolling}, we specifically used 165 images from the GlaS dataset, all of which were adjusted to the dimensions of $512 \times 512$.

\paragraph{CVC-ClinicDB}
The CVC-ClinicDB dataset~\cite{bernal2015wm}, often abbreviated simply as "CVC," serves as a publicly accessible resource for polyp diagnosis within colonoscopy videos. It encompasses a total of 612 images, each having a resolution of $384 \times 288$, meticulously extracted from 31 distinct colonoscopy sequences. These frames provide a diverse array of polyp instances, making them particularly useful for the development and evaluation of polyp detection algorithms. To ensure consistency across different datasets used in our study, all images from the CVC-ClinicDB dataset were uniformly resized to $256 \times 256$.

\subsection{Implementation Details}
\paragraph{Segmentation U-KAN} We implemented \model~using Pytorch on a NVIDIA RTX 4090 GPU. For the BUSI, GlaS and CVC datasets, the batch size was set to 8 and the learning rate was 1e-4.
  We used the Adam optimizer to train the model, and used a cosine annealing learning rate scheduler with a minimum learning rate of 1e-5. The loss function was a combination of binary cross entropy (BCE) and dice loss. We randomly split each dataset into $80 \%$ training and $20 \%$ validation subsets. 
All the results among these datasets are reported over three random runs.
 Only vanilla data augmentations including random rotation and flipping is applied. We trained the model for 400 epochs in total.
 We compare the output segmentation images both qualitatively and quantitatively using various metrics such as IoU and F1 Score.
We also report the metrics related to computation cost such as Gflops and number of parameters (Params).

\paragraph{Diffusion U-KAN} 
The image was cropped and resized into $64\times64$ for unconditional generation. We benchmark all the methods with the same training setting: 1e-4 learning rate, 1000 epochs, Adam optimizer, and cosine annealing learning rate scheduler. 
To evaluate the generation capacity of each method, we generate 2048 image samples using random Gaussian noise as input. We then compare the generated images qualitatively and quantitatively using various metrics such as Fréchet Inception Distance (FID)~\cite{parmar2021buggy} and Inception Score (IS)~\cite{saito2017temporal}. These metrics provide insights into the diversity and quality of the generated images.

\subsection{Performance Comparison on Image Segmentation}

Tab.~\ref{tab:exp_seg} presents the results of the proposed \model~against all the compared methods over all the benchmarking datasets.
Comparisons between our \model~and recently favored frameworks for medical image segmentation were conducted, benchmarking against convolutional baseline models such as U-Net~\cite{ronneberger2015u}, U-Net++~\cite{zhou2018unet++}. We also evaluated performance against attention-based counterparts including Att-UNet~\cite{oktay2018attention} and the state-of-the-art efficient transformer variant, U-Mamba~\cite{ma2024u}.
Furthermore, as KAN emerges as a promising alternative of MLP, we further perform comparison against the advanced MLP-based segmentation networks, including U-NeXt~\cite{valanarasu2022unext} and Rolling-UNet~\cite{liu2024rolling}.
In terms of the performance metrics, two standard metrics including Intersection over Union (IoU) and F1 scores are used for evaluating image segmentation tasks.
The results demonstrate that across all datasets, our \model~ surpasses the performance of all other methodologies.

\begin{table*}[htbp]
  \centering
  \makeatletter\def\@captype{table}\makeatother
  \caption{
Comparison with state-of-the-art segmentation models on three heterogeneous medical scenarios. The average results with standard deviation over three random runs are reported.
  }
  \resizebox{\linewidth}{!}{
  
   \setlength{\tabcolsep}{1.8mm}
    \begin{tabular}{lcccccc}
    \toprule
    \multirow{2}[4]{*}{Methods} & \multicolumn{2}{c}{BUSI~\cite{al2020dataset}}    & \multicolumn{2}{c}{GlaS~\cite{valanarasu2021medical}}    & \multicolumn{2}{c}{CVC~\cite{bernal2015wm}} \\
\cmidrule{2-7}                 & IoU↑         & F1↑          & IoU↑         & F1↑          & IoU↑         & F1↑ \\
    \midrule
    U-Net~\cite{ronneberger2015u}      &  57.22±4.74 & 71.91±3.54 & 86.66±0.91 & 92.79±0.56 & 83.79±0.77 & 91.06±0.47\\
    Att-Unet~\cite{oktay2018attention}     & 55.18±3.61   & 70.22±2.88   & 86.84±1.19   & 92.89±0.65   & 84.52±0.51 & 91.46±0.25
 \\
    U-Net++~\cite{zhou2018unet++}       & 57.41±4.77   & 72.11±3.90   & 87.07±0.76   & 92.96±0.44   & 84.61±1.47
  & 91.53±0.88
 \\
    U-NeXt~\cite{valanarasu2022unext}        & 59.06±1.03   & 73.08±1.32   & 84.51±0.37   & 91.55±0.23   &74.83±0.24 & 85.36±0.17
 \\
    Rolling-UNet~\cite{liu2024rolling}    & 61.00±0.64   & 74.67±1.24   & 86.42±0.96   & 92.63±0.62   & 82.87±1.42   & 90.48±0.83
 \\
    U-Mamba~\cite{ma2024u}      &  61.81±3.24 & 75.55±3.01 & 87.01±0.39 & 93.02±0.24 & 84.79±0.58 & 91.63±0.39 \\
\midrule
   \rowcolor{pink!12} Seg. \model~(Ours) &   \textbf{  63.38±2.83} & \textbf{76.40±2.90} & \textbf{87.64±0.32} & \textbf{93.37±0.16} & \textbf{85.05±0.53} & \textbf{91.88±0.29}       \\
    \bottomrule
    \end{tabular}%
    
    }
  \label{tab:exp_seg}%
\end{table*}%

In addition to the accuracy benefits, this paper further demonstrates the efficiency of our method when used as a network baseline. As shown in Tab.~\ref{tab:exp_efficiency}, we report the model's parameter volume (M) and Gflops on various datasets, as well as segmentation accuracy. The results indicate that our method not only surpasses most segmentation methods in terms of segmentation accuracy, but also exhibits significant advantages or comparable levels in terms of efficiency, with the exception of UNext. Overall, in the trade-off between segmentation accuracy and efficiency, our method exhibits the best performance.

\begin{table*}[htbp]
\centering
\caption{
Overall comparison with state-of-the-art segmentation models w.r.t. efficiency and segmentation metrics.
}
\setlength{\tabcolsep}{3.5mm}
\begin{tabular}{lcccc}
\toprule
\multicolumn{1}{l}{\multirow{2}[4]{*}{Methods}} & \multicolumn{2}{c}{Average Seg.} & \multicolumn{2}{c}{Efficiency} \\
\cmidrule{2-5} & IoU↑ & F1↑ & Gflops & Params\,(M) \\
\midrule
U-Net~\cite{ronneberger2015u} & 75.89±2.14
 &85.25±1.52
 & 524.2
& 34.53
\\
Att-Unet~\cite{oktay2018attention} & 75.51±1.77 & 84.85±1.26 & 533.1 & 34.9 \\
U-Net++~\cite{zhou2018unet++} & 76.36±2.33 & 85.53±1.74 & 1109 & 36.6 \\
U-NeXt~\cite{valanarasu2022unext} & 72.80±0.54 & 83.33±0.57 &  4.58 & {1.47} \\
Rolling-UNet~\cite{liu2024rolling} & 76.76±1.01 & 85.92±0.89 & 16.82 & {1.78} \\
U-Mamba~\cite{ma2024u} & 77.87±1.47 & 86.73±1.25 & 2087 & 86.3 \\
\midrule
 \rowcolor{pink!12}Seg. \model~(Ours) & \textbf{78.69±1.27} & \textbf{87.22±1.15} &{14.02} & 6.35\\
\bottomrule
\end{tabular}%
\label{tab:exp_efficiency}%
\end{table*}%

\begin{figure*}[!t]
	\begin{center}
		\includegraphics[width=0.95\linewidth]{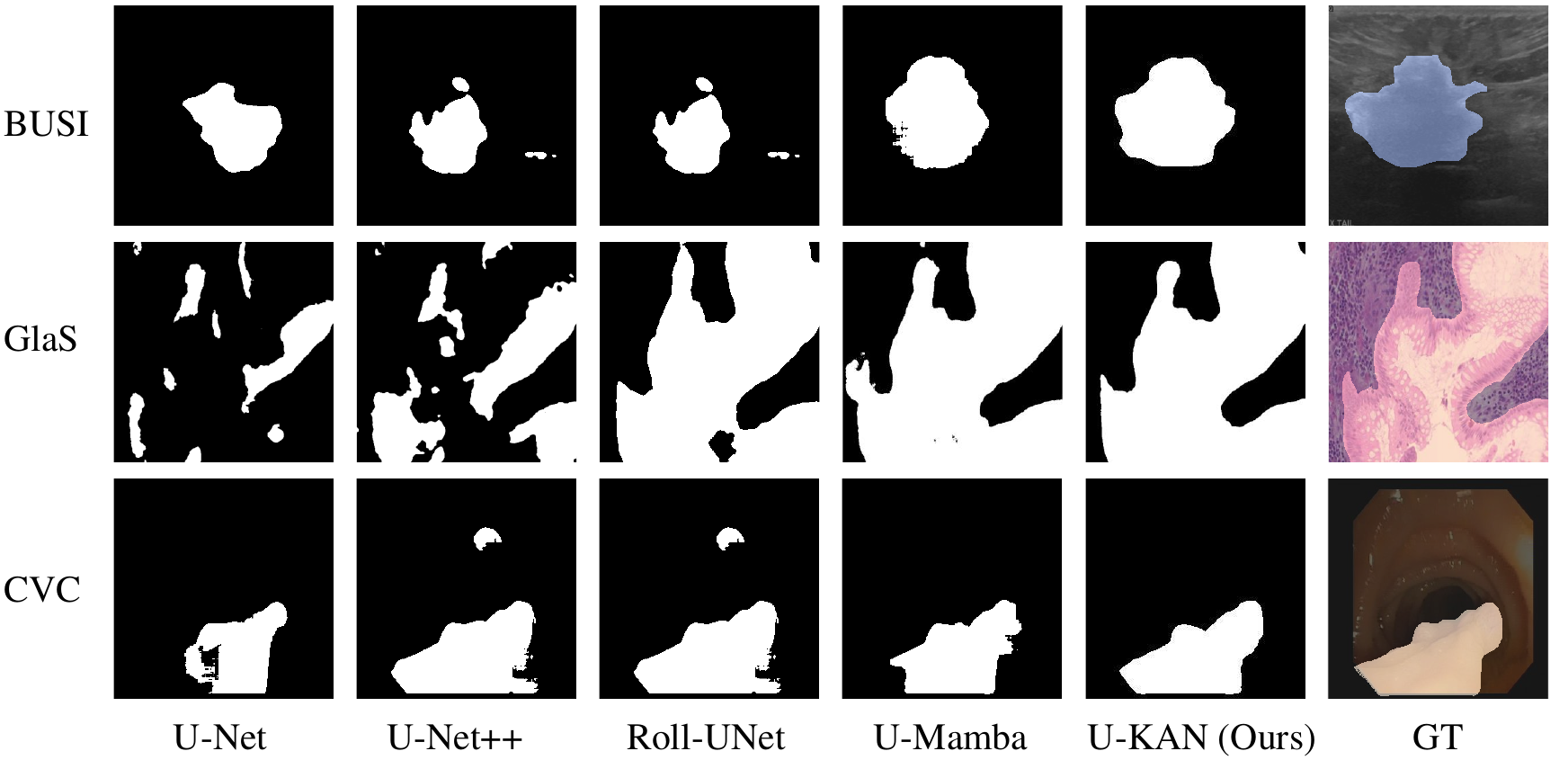}
	\end{center}
	\caption{
 Visualized segmentation results of the proposed \model~against other state-of-the-arts over three heterogeneous medical scenarios.
 }
	\label{fig:exp_seg}
\end{figure*}

We further present a comprehensive qualitative comparison across all datasets, as depicted in Fig. \ref{fig:exp_seg}. Firstly, it is evident from the results that pure CNN-based approaches such as U-Net and U-Net++ are more prone to over- or under-segmentation of organs, suggesting the limitation of these models in encoding global context and discriminating semantics. 
In contrast, our proposed \model~yields fewer false positives compared to other methods, indicating its superiority in suppressing noisy predictions. 
When juxtaposed with models based on Transformers, and efficient MLP-based architectures, the predictions of \model~often exhibit finer details in terms of boundaries and shapes. These observations underscore \model's capability for refined segmentation while preserving intricate shape information.
This further corroborates our initial intuition, highlighting the advantages introduced by incorporating the KAN layer.

\subsection{Performance Comparison on Image Generation}
We investigated the potential of our proposed \model~as a backbone for genertive tasks. We compared our \model~with various diffusion variant models, all based on conventional U-Nets, in order to evaluate the efficacy of this architecture for different generative tasks. The results were presented in Tab.~\ref{tab:exp_gen}, where we reported FID~\cite{parmar2021buggy} (Fréchet Inception Distance) and IS~\cite{saito2017temporal} (Inception Score) metrics across three datasets.
The Fréchet Inception Distance (FID) measures the distance between distributions of generated and real images. Lower FID indicates better resemblance to real images. The Inception Score (IS) evaluates image quality by classification accuracy into categories, with higher IS indicating better classification.
The results from our experiments clearly indicate that our method exhibits superior generative performance compared to other state-of-the-art models in the field. This suggests that the architecture of our \model~is particularly suitable for generative tasks, providing an effective and efficient approach to generating high-quality images.

\begin{table*}[htbp]
  \centering
  \small
  \makeatletter\def\@captype{table}\makeatother
  \caption{Comparison with standard U-Net based diffusion models on three heterogeneous medical scenarios. Results by different variants of Diffusion U-Net is provided for comprehensive evaluation.}
   \setlength{\tabcolsep}{1.0mm}
    \begin{tabular}{llcccccc}
    \toprule
\multirow{2}[4]{*}{Methods} &\multirow{2}[4]{*}{Middle Blocks} & \multicolumn{2}{c}{BUSI~\cite{al2020dataset}}    & \multicolumn{2}{c}{GlaS~\cite{valanarasu2021medical}} & \multicolumn{2}{c}{CVC~\cite{bernal2015wm}} \\
\cmidrule{3-8} & & FID↓ & IS↑ & FID↓ & IS↑ & FID↓ & IS↑ \\
\midrule
\multirow{3}{*}{Diffusion U-Net} &ResBlock+Attn & 116.52 &2.54 & 42.65 &{2.45} & 49.30 & 2.65 \\
&Identity & 124.46 & 2.71 & 42.63 &2.41 & 50.42 & 2.49 \\
&MLP & 104.95 & 2.59 & 44.21 &2.43 & 51.16 & 2.69 \\
\midrule
\rowcolor{pink!12} Diffusion U-KAN~(Ours)&KANBlock & \textbf{101.93} & \textbf{2.76} & \textbf{41.55} & \textbf{2.46} & \textbf{46.34} & \textbf{2.75} \\
    \bottomrule
    \end{tabular}%
  \label{tab:exp_gen}%
\end{table*}%

\begin{figure*}[!t]
	\begin{center}
		\includegraphics[width=0.88\linewidth]{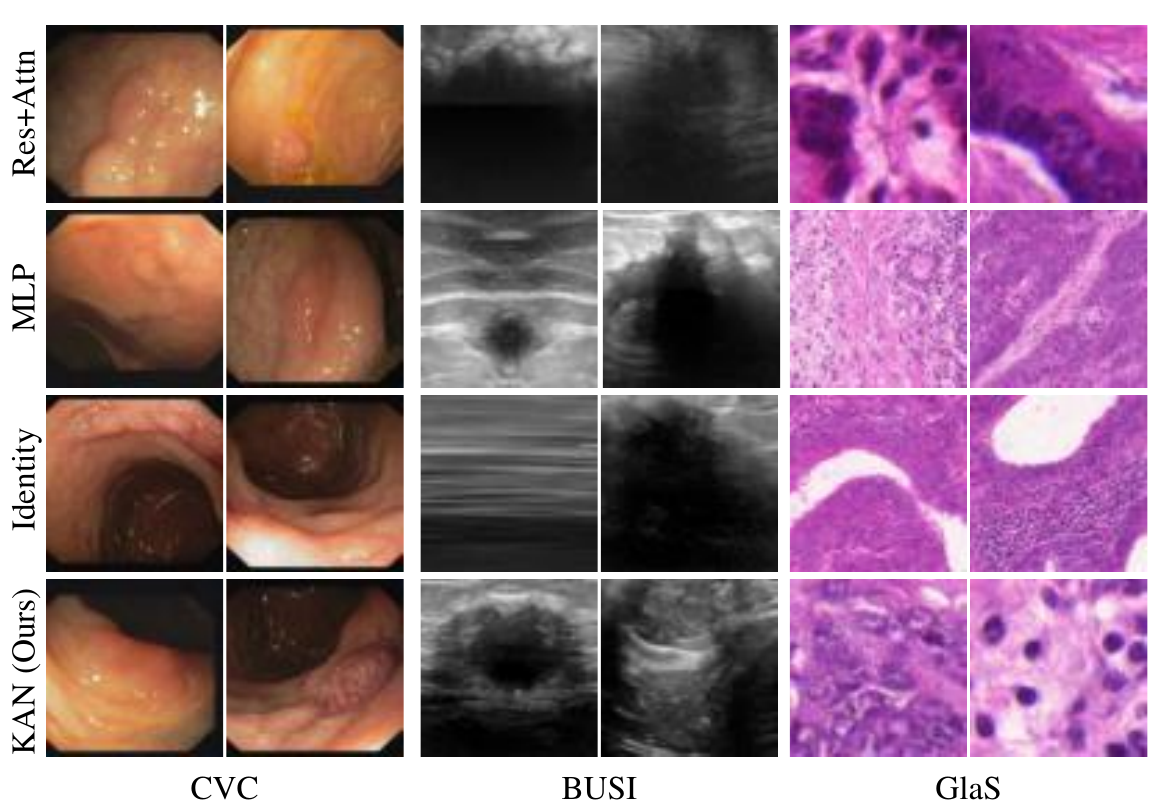}
	\end{center}
	\caption{Generated images by proposed Diffusion U-KAN in three heterogeneous medical scenarios.}
	\label{fig:exp_gen}
\end{figure*}

Fig.~\ref{fig:exp_gen} displays visualizations of some of our generated results. It is observed that our method can produce realistic and diverse content across multiple distinct datasets, demonstrating its versatility and effectiveness in generating high-quality images. This further supports the claim that \model~has a significant advantage when it comes to generative tasks, making it a strong candidate for future research and development in this area.

\subsection{Ablation Studies}
To thoroughly evaluate the proposed TransUNet framework and validate the performance under different settings, a variety of ablation studies were performed as follows.
=

\begin{table*}[h]
\centering
\begin{minipage}{0.48\textwidth}
    \centering
    \makeatletter\def\@captype{table}\makeatother
    \caption{
Ablation studies on number of used KAN layers. 
The {\colorbox{pink!12}{default} setup is {denoted}}.
    }
    \setlength{\tabcolsep}{1.6mm}
    \label{tab:reorder}
	\centering
	\vspace{1pt}
\begin{tabular}{l|ccc}
\toprule
\#KAN & IoU↑ & F1↑ & Gflops \\
\midrule
1 Layer & 64.20 & 77.81 & 13.97 \\
2 Layer& 64.56 & 78.01 & 14.00 \\
 \rowcolor{pink!12}3 Layer& 65.26 & 78.75 & 14.02 \\
4 Layer& 64.72 & 78.35 & 14.05 \\
5 Layer& 64.86 & 78.42 & 14.07 \\
\bottomrule
\end{tabular}
 \label{tab:exp_abl1}
\end{minipage}
\hfill
\begin{minipage}{0.48\textwidth}
\centering
    \makeatletter\def\@captype{table}\makeatother
    \caption{
    Ablation studies on using KAN layers against MLPs. 
    The {\colorbox{pink!12}{default} setup is {denoted}}.
    }
    	\vspace{1pt}
    \setlength{\tabcolsep}{1.6mm}
    \label{tab:reorder}
	\centering
\begin{tabular}{l|ccc}
\toprule
KAN vs. MLP & IoU↑ & F1↑ & Gflops \\
\midrule
 \rowcolor{pink!12}KAN$\times$3 & 65.26 & 78.75 & 14.02 \\
MLP+KAN+KAN & 64.12 & 77.86 & 14.29 \\
KAN+MLP+KAN & 63.82 & 77.58 & 14.29 \\
KAN+KAN+MLP & 64.30 & 77.95 & 14.29 \\
MLP$\times$3 & 63.49 & 77.07 & 14.84 \\
\bottomrule
\end{tabular}
  \label{tab:exp_abl2}
  
\end{minipage}
\end{table*}

\begin{table*}[!t]
  \centering
  \caption{
  Ablation studies on model scaling by using different channel settings in \model. The {\colorbox{pink!12}{default} setup is {denoted}}.
  }
     \setlength{\tabcolsep}{3mm}
    \begin{tabular}{l|cccccc}
    \toprule
    Model Scale & $C_1$ & $C_2$ & $C_3$ & IoU↑ & F1↑ & Gflops \\
\midrule
\model-S &64&96&128 & 64.62 & 78.28 &3.740\\
 \rowcolor{pink!12} \model &128&160&256  & 65.26 & 78.75 & 14.02  \\
\model-L&256&320&512  & 66.01 & 79.09 & 55.11 \\
    \bottomrule
    \end{tabular}%
  \label{tab:exp_abl3}%
\end{table*}%

\paragraph{The Number of KAN Layer}
As previously stated, the inclusion of KAN Layers in~\model~has proven beneficial by facilitating the modeling of more refined segmentation details through the explicit incorporation of highly efficient embeddings. The objective of this ablation study was to assess the impact of incorporating varying quantities of KAN Layers.
We modified the number of KAN Layers from one to five, as depicted in Tab.~\ref{tab:exp_abl1}.
It is observed that the configuration with three KAN Layers yielded the most superior performance.
These outcomes demonstrate that the strategic integration of an adequate number of KAN Layers within the~\model~can effectively capture intricate segmentation-related nuances.

\paragraph{Impact on Using KAN Layer v.s. MLP}
To further substantiate the role of KAN layers in enhancing model performance, we conducted an array of ablation experiments, as shown in Tab.~\ref{tab:exp_abl2}. In these experiments, we replaced the introduced KAN layers with traditional multilayer perceptrons (MLPs) to observe if such modifications would result in a decrease in performance. This methodology allowed us to more tangibly comprehend the significance of KAN layers in improving the model's overall performance.
Initially, we modified a model that already incorporated KAN layers, replacing one or several KAN layers with standard MLPs. Subsequently, using identical datasets and training parameters, we retrained the modified model and documented its performance across various tasks.
The outcomes demonstrated a noticeable decline in performance across multiple tasks when KAN layers were replaced with MLPs, particularly in intricate tasks requiring robust feature extraction and representational capacities. These findings underscore the crucial role of KAN layers in augmenting the model's expressive capabilities and bolstering its overall performance.

\paragraph{Model Scaling}
We conducted an ablation study on various model sizes of the \model. Specifically, we examined alternative configurations of the \model, termed as \textit{Small} and \textit{Large} models. The primary distinction between these variants lies in their channel settings, denoted as the varied channel number from first to third KAN layer ($C_1$-$C_3$), as detailed in Tab. \ref{tab:exp_abl3}. The \textit{Small} model features channel settings of 64-96-128, while the \textit{Large} model's channel counts are set to 256-320-512. In contrast, our default model's channel numbers are configured at 128-160-256. We observed that larger models correlate with enhanced performance, which aligns with the scaling law characteristics exhibited by models integrating KAN. Ultimately, to strike a balance between performance and computational expenses, we opted to employ the default base model in our experiments.

\begin{figure}[!t]
	\begin{center}
		\includegraphics[width=\linewidth]{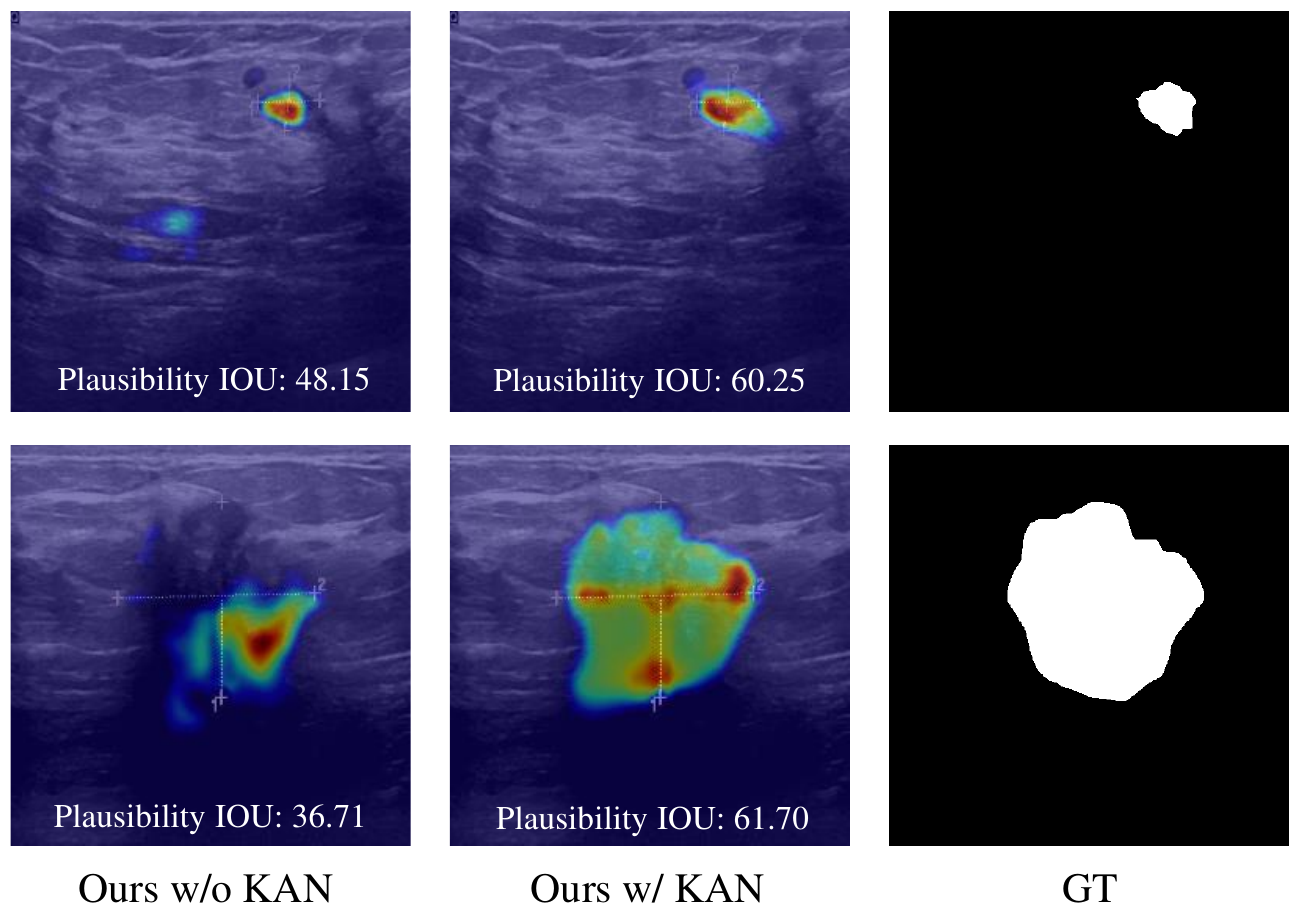}
	\end{center}
 \vspace{-1em}
	\caption{Explainability of U-KAN with channel activation.}
	\label{fig:explain}
\end{figure}

\paragraph{Explainability}
We further explore the interpretability of KAN layers by analyzing activated patterns, as depicted in Fig. \ref{fig:explain}. When utilizing MLP layers (1th column), the model struggles to identify appropriate activation regions essential with an unsatisfactory Plausibility IoU, which is a metric provided in~\cite{cambrin2024kanitkanssentinel} that calculates IoU between thresholded activation maps and GT masks (higher is better).
In contrast, with integrating KAN layer (2nd column), there is a marked improvement in the ability to precisely locate the region of interest and activate the boundaries that align closely with the ground truth (3rd column). {This underscores the pivotal role of KAN layers in enhancing the explainable decision-making of deep models}, especially for mask prediction, which is also aligned with the observation in KAN~\cite{liu2024kan}.

\section{Conclusion}
This paper introduces \model~and demonstrates the significant potential of Kolmogorov-Arnold Networks (KANs) in enhancing backbones like U-Net for various visual applications. By integrating KAN layers into the U-Net architecture, you can make a strong network for vision tasks in terms of impressive accuracy, efficiency and interpretability. 
We perform empirical evaluations of our method under several medical image segmentation tasks.
Moreover, the adaptability and effectiveness of \model~also highlight its potential as a superior alternative to U-Net for noise prediction in diffusion models. 
These findings underscore the importance of exploring non-traditional network structures like KANs for advancing a broader range of vision applications.

\paragraph{Future Work}
Future endeavors will involve extending these advanced network operators to more extensive of settings and higher-dimensional data formats, such as temporal data~\cite{genet2024tkan,wang2020deep}, genomic data~\cite{waqas2024senmo,poirion2021deepprog,li2024gtp} and 3D representations~\cite{moryossef2024optimizing,mildenhall2021nerf,pan2023learning,li2023steganerf,li2024gaussianstego}.

\appendix

\bibliography{aaai25}

\end{document}